\documentclass{aa}  
\usepackage{graphicx}
\usepackage{txfonts}
\usepackage{amssymb}
\usepackage{amsfonts}

\newcommand\T{\rule{0pt}{2.6ex}}
\newcommand\TT{\rule{0pt}{3.0ex}}
\newcommand\B{\rule[-1.2ex]{0pt}{0pt}}
\newcommand\BB{\rule[-1.4ex]{0pt}{0pt}}

\begin{document}
   \title{GRS 1915+105 : High-energy Insights with \textit{SPI/INTEGRAL}}

   \subtitle{Spectral Analysis of the Comptonized Emission}

   \author{R. Droulans
          \and
          E. Jourdain
	  }

   \offprints{R. Droulans}

   \institute{CESR / CNRS -- Universite de Toulouse, 9 Av. du Colonel Roche, 31028 Toulouse Cedex~04, France}

   \date{Received ; accepted}

   \titlerunning{SPI/INTEGRAL views on GRS 1915+105}
   \authorrunning{R. Droulans \& E. Jourdain}

 \abstract
   {We report on results of nearly two years of INTEGRAL/SPI monitoring of the
Galactic microquasar GRS 1915+105.}
   {From September 2004 to May 2006, the source
has been observed twenty times with long ($\sim $100 ks) exposures. We present an
analysis of the SPI data and focus on the description of the high-energy
($>$\thinspace 20 keV) output of the source.}
   {Temporal and spectral analysis of the SPI data. Comparison to simultaneous 1.2 -- 12 keV ASM data.}
   {We found that the 20\thinspace --\thinspace 500 keV
spectral emission of GRS 1915+105 was bound between two states. It seems that these high-energy states
are not correlated with the temporal behavior of the source, suggesting that there is no
direct link between the macroscopic characteristics of the coronal plasma and the the variability of the accretion flow.
All spectra are well fitted by a thermal comptonization component plus an extra
high-energy powerlaw. This confirms the presence of thermal and non-thermal
electrons around the black hole.}
   {}


   \maketitle
%

\section{Introduction}

GRS 1915+105 is among the most notorious accreting black holes in our
Galaxy. Not only is it one of the brightest and most variable X-ray sources
in the sky (see Castro-Tirado et al. 1992 for a first detection report and
Belloni et al. 2000 for detailed variability analysis), but it is also the
first Galactic object in which superluminal plasma ejections have been
observed (Mirabel \& Rodriguez 1994). Consequently, much work has been done to
try to understand the complex nature of the accretion flow (see e.g. the
comprehensive analysis by Done, Wardzi\'nski \& Gierli\'nski 2004) and the
disk-jet coupling in this binary system. An extensive review on these
aspects can be found in Fender \& Belloni (2004). Very recently, Rodriguez
et al. (2008a \& 2008b) presented an extended report on GRS 1915+105 discussing insights gained from a $2$-year long
monitoring campaign involving multi-wavelength observations.

However, most of the recent studies focus on the astonishing X-ray properties, the soft $%
\gamma $-ray emission ($>$\thinspace 100 keV) being generally observed at poor signal to noise
ratio. Understanding the high-energy behavior of the source is nevertheless very
important as it is assumed to trace the physical processes occurring at the
innermost regions surrounding the black-hole (e.g. Galeev et al. 1979 or Malzac 2007 for a review). 
Featuring good spectral resolution and sensitivity up to several $%
MeV$, SPI is a good instrument to tackle this issue.

The emission ranging from hundreds of $eV$ up to almost $1\,MeV$ is
generally assumed to originate from two components (e.g. Vilhu et al 2001). 
While the softer part is unanimously attributed
to thermal emission from the disk, various
interpretations exist concerning the origin of the high-energy emission. Detailed analysis
of joint RXTE/OSSE observations allowed Zdziarski et al. (2001 \& 2005) to
rule out some of these interpretations. They suggested that the
observed $10$\thinspace --\thinspace $500\,keV$ emission originated from
inverse Compton scattering of the soft disk photons by a hybrid
thermal/non-thermal electron population.

In this paper we present all SPI observations of GRS 1915+105 from September 2004
to May 2006. We focus on gaining the most accurate high-energy picture of the source,
mainly through extensive spectral analysis. First we briefly
describe the instrument as well as our data reduction methods. The
observational results are presented and further
discussed in the concluding part of the paper.

\section{Observations and data reduction}

\subsection{INTEGRAL/SPI}

SPI is a high resolution $\gamma $-ray spectrometer (Vedrenne et al. 2003)
aboard the INTEGRAL observatory. The observational strategy of the
INTEGRAL mission is based on approximately $3$-day long revolutions during
which one or several fields of view are sampled by means of $30$\thinspace --\thinspace $%
40$ minute long fixed pointings, separated by a $2{{}^\circ}$ angular distance. As the number of SPI pixels is small (among
the initial $19$ detectors, two broke down reducing the detector plane to only $17$
pixels), this so-called \textquoteleft dithering\textquoteright\
scheme (see Jensen et al. 2003 for details) is essential for SPI image
reconstruction. In particular, combining the data from a set of dithered
pointings allows one to enhance the precision of flux extraction (see e.g. Joinet et al. 2005).
 
\subsection{Data analysis}

We analysed all public data from nearly two years of SPI observations on GRS
1915+105 (September 2004 -- May 2006). For each revolution, we first gathered all
pointings where GRS 1915+105 was less than $12{{}^\circ}$ off the central axis. Pointings showing contamination by solar flares or
radiation belt exit/entry were excluded. The log of the
resulting $1.7\,Ms$ of observational coverage is given in Table 1. In order
to detect the emitting sources in the field of view, we used the SPIROS
software (Skinner \& Connell 2003) to produce $20$\thinspace --\thinspace $%
50\,keV$ images for each observation. The positions of the active sources
(detected at a minimum of $5\,\sigma $) were then given as input to a specific
flux-extraction algorithm, using the SPI instrument response for sky-model fitting. From there we started to analyse the background and
source behavior for each observation. We first allowed the background
pattern as well as the main source(s) to vary between successive pointings.
In this way we determine each component's most appropriate variability
timescale for the final data reduction. We used pointing-durations ($\approx 2\,ks$) as a
timescale for the GRS 1915+105 light-curves and adapted the time sampling
for subsequent extraction of the $20\,keV$\thinspace --\thinspace $8\,MeV$ spectra
according to the observed temporal behavior. This method allowed us to minimize
the error bars without losing any scientific information.

\section{Results}

\subsection{Light-curves}

Figure 1 displays the total $20$\thinspace --\thinspace $50\,keV$ light-curve for our observational period.
GRS 1915+105 shows relevant long-term variability, the averaged source flux per observation
($\approx$ one-day) spans between $90$ and $380\,mCrab$, with an approximate uncertainty of $5\,mCrab$.
\begin{figure}[ht]
  \begin{center}
  \includegraphics[angle=90,width=\columnwidth]{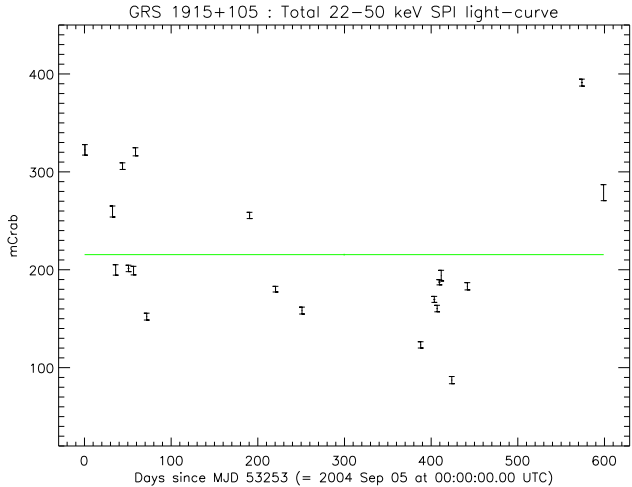}
     \caption{Long-term variability of GRS 1915+105 in the SPI band. Each point represents the mean $20$\thinspace --\thinspace $50\,keV$
	flux per observation (corresponding to a timescale of about one-day, see Table 1).}
     \label{lctotal}
     \end{center}
\end{figure}
On the shorter science-window timescale ($\approx$\thinspace $2\,ks$), the obtained individual light-curves show in general lower variability.
Within a single observation, the source flux varies by at most a factor of $2$. 
As GRS 1915+105 is well known for being very variable in X-rays, we also
considered SPI-simultaneous $1.2$\thinspace --\thinspace $12$ keV ASM
light-curves to compare the X- and soft $\gamma $-ray
behavior of the source. We found an
anti-correlation trend between the $\sim$\thinspace one-day averaged $1.2$\thinspace --\thinspace $12$ and $%
20 $\thinspace --\thinspace $50\,keV$ source fluxes (Figure 2) 
with a linear correlation factor $\rho$ of $-0.59 \pm 0.02$ and a $99\%$ significance. 
On the science-window timescale however, a positive flux-correlation
is discernible for observation 368, as
the source exhibits very strong variability (Figure 5 right).
\begin{figure}[ht]
  \begin{center}
  \includegraphics[angle=90,width=\columnwidth]{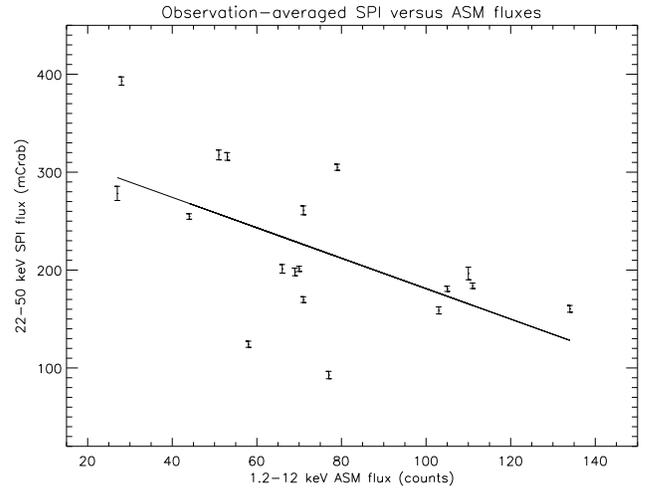}
     \caption{One day timescale flux-flux relation between the $1.2$\thinspace --\thinspace $12$ and $%
	20 $\thinspace --\thinspace $50\,keV$ bands. We observe an anticorrelation in the flux trend between the two bands.}
     \label{SPI_ASM}
     \end{center}
\end{figure}
The variability amplitudes (parameterized by fractional rms) are not straightforward to compare since the time bins are very 
different between the two instruments.
More importantly, the $1\sigma$ SPI errors for the individual light-curves are of the same order of magnitude than absolute rms variability, 
thus limiting a solid scientific interpretation.

\subsubsection{Spectral analysis}

We used XSPEC 11.3.2 (Arnaud 1996) for spectral analysis. First, each observation-averaged spectrum
was fitted with a basic powerlaw model which gives a good
description of most of the data. Then we characterized the spectra by $20 $\thinspace --\thinspace $50\,keV$ flux and
best-fit photon index, the latter being found to range from $2.8$ to $3.5$, with typical uncertainty of $0.1$. 
The results are summarized in Table 1 and
illustrated by a photon index versus flux diagram given in Figure 3. We used this
figure to select four observations which will be presented in
more detail in the following subsections. 
\begin{figure}[ht]
   \centering
   \includegraphics[angle=90,width=\columnwidth]{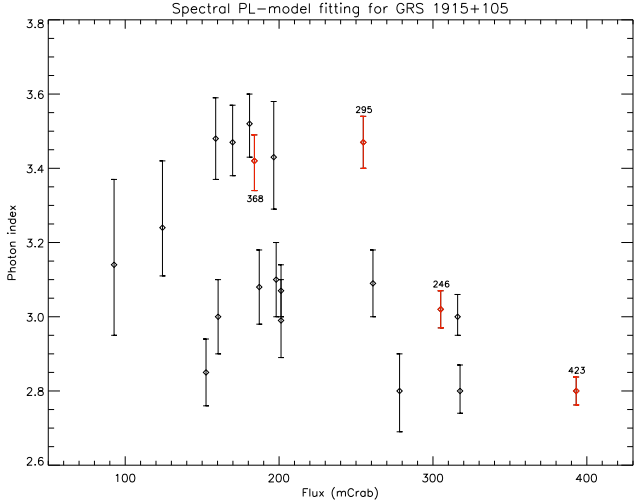}
      \caption{Basic source characteristics : model photon index versus mean  flux per observation. 
	The observations that are discussed in the text are highlighted in red.}
      \label{indflux}
\end{figure}
We selected the observations made during INTEGRAL revolutions 295 and 423 since they 
both have rather high average fluxes (and therefore good signal to noise ratios), but at the same time very different spectral shapes.
Further, the light-curves showed no significant variability during these observations
(neither in the $20$\thinspace --\thinspace $50\,keV$ band nor in X-rays),
hence allowing a meaningful analysis of the observation-averaged spectrum. In addition, we selected two other
observations (Rev. 246 and 368) because they were part of large multi-wavelength campaigns, allowing us to put the SPI data in a broader context.

\subsubsection{Observations 295 and 423}

From the ASM light-curves in Figure 4, we see that during both observations GRS
1915+105 showed very similar X-ray activity, characterized by
low flux and almost no variability. The temporal properties in the 
$20$\thinspace --\thinspace $50\,keV$ band are also quite similar, 
with again little variability during both observations.
On the other hand, spectral characteristics are found to be significantly different, 
with photon indices of $3.47 \pm 0.07$ and $2.80 \pm 0.04$ respectively. 
We then investigated these
differences through more detailed spectral modelling.
\begin{figure}[!t]
   \centering
   \includegraphics[angle=90,width=\columnwidth]{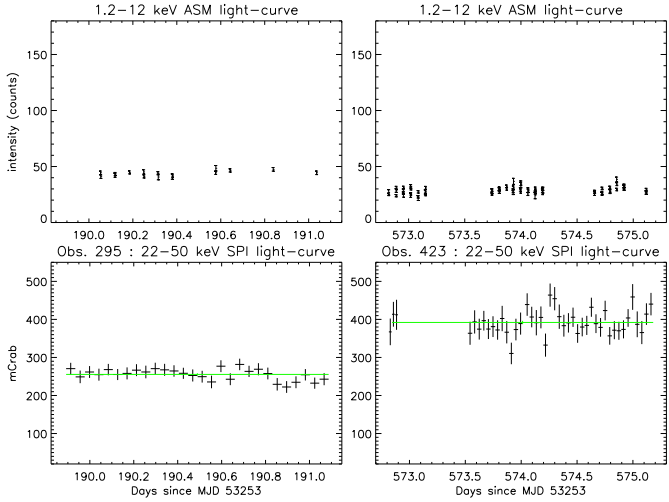}
      \caption{ASM and SPI light-curves for observations 295 and 423. For both observations the source exhibits low flux and very little variability
	in the ASM band. In the $20$\thinspace --\thinspace $50\,keV$ SPI band we observe similar stability on the science-window timescale
	with intermediate and high average flux respectively.}
      \label{lc2954236}
\end{figure}
\begin{figure}[!t]
   \centering
   \includegraphics[angle=90,width=\columnwidth]{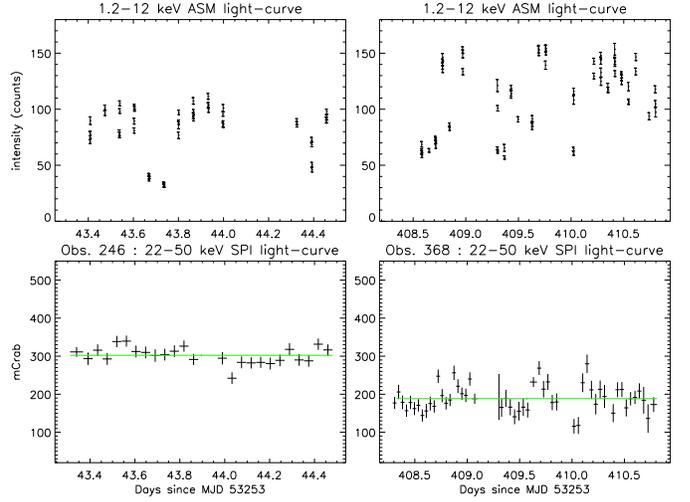}
      \caption{ASM and SPI light-curves for observations 246 and 368. During the latter a correlation 
        (on the $\approx$\thinspace $2\,ks$ science-window timescale) between 
	the $1.2$\thinspace --\thinspace $12$ and $20$\thinspace --\thinspace $50\,keV$ bands is noticeable, hinting that both bands are tracing the 
	evolution of the comptonized emission.}
      \label{lc368246}
\end{figure}
From Table 1 one can see that the simple powerlaw model gives a rather poor fit
for observation 295. More precisely, we noticed that above $100\,keV$ all
spectral points are located above the fitted powerlaw. Pure thermal
comptonization models like COMPTT (where the comptonized spectrum is completely determined by the plasma temperature and its optical 
depth, Titarchuk 1994) can thus be ruled out ($\chi{{}^2}/\nu =72/26$, see table 2). 
Assuming that the low-energy part is nonetheless produced
through thermal comptonization, one needs to add a further spectral
component. We thus chose to add a powerlaw as a phenomenological description of the observed high-energy tail.
Given the error amplitude above $100\,keV$, we arbitrarily fixed the photon 
index to $2.0$.
The resulting fit ($kT_{e}={16.3}_{-0.9}^{+1.2}\,keV$ and $\tau=0.57$\footnote{fixed to its best-fit value for error 
calculation on $kT_{e}$}) is in very good agreement with our 
data ($\chi{{}^2}/\nu =17/25$) and the \textit{F-TEST} indicates a probability 
of $\approx$\thinspace $10^{-9}$ that this improvement has been a chance event. As a last step we 
applied the Poutanen \& Svensson (1996) COMPPS model
which describes comptonization from a hybrid thermal/non-thermal electron plasma. This iterative scattering method assumes 
a powerlaw distribution ($\propto \gamma^{-\Gamma_{e}}$) up to a Lorentz factor $\gamma_{min}$ below 
which the plasma thermalizes to a Maxwellian 
distribution. We fixed a spherical geometry and assumed the reflection component to be insignificant, enabling a 
straightforward comparison with COMPTT+PL.
Note that COMPPS poses more constraints than COMPTT+PL, as it requires both 
components (thermal and non-thermal) to be linked, whereas 
the former does not. Equally good fitting results ($\chi{{}^2}/\nu =17/25$) for 
both models indicate that the observed  
$20$\thinspace --\thinspace $500\,keV$ emission is most likely to originate from thermal 
and non-thermal comptonization processes.

For observation 423, the basic powerlaw fit is clearly unacceptable ($\chi {{}^2}/\nu =83/27$) due to the marked curvature around $50\,keV$, leaving clear
evidence for thermal processes. Replacing the powerlaw by a pure thermal
component improved the fit and gave $\chi {{}^2}/\nu =42/26$. However, this interpretation is not able to account for the observed
emission above $200\,keV$, again suggesting the presence of an additional
component. We thus keep the hybrid comptonization models ($\chi{{}^2}/\nu =37/25$ for COMPTT+PL and $\chi{{}^2}/\nu =37/24$ for COMPPS) as our preferred 
description for observation 423. 

All fitting results are summarized in Table 2 and discussed in section 4.

\subsection{Observations 246 and 368}

Both observations have already been intensively studied. Observation 246 is
discussed by Rodriguez et al. (2008b) who conducted detailed spectro-temporal
analysis by means of multi-wavelength coverage. They found that the
source showed periodic X-ray cycles, identified as alternations of $\nu $
and $\rho $ classes (see Belloni et al. 2000 for definitions). 
INTEGRAL observation 368 was again
part of a large multi-wavelength campaign, this time involving the Suzaku
satellite (Ueda et al. 2006). These authors identified the observed X-ray
variability pattern to be the signature of a high flux transition from a soft
class $\chi $ to class $\theta $. We find the same variability pattern
in X-rays as in the $20$\thinspace --\thinspace $50\,keV$ SPI band (Figure 5), indicating that both bands are probably sampling the temporal behavior of the same component.
For observation 246 there seems to be a similar correspondence, although less clear due to lower variability amplitude in the SPI band.
Both spectra are fitted with thermal\thinspace +\thinspace non-thermal comptonization models (COMPTT+PL and COMPPS)
which provide the best
agreement to the data (see Table 2). Compared to the joint JEMX/ISGRI spectrum
from observation 246 (Rodriguez et al. 2008b), our SPI data are found to
have almost the same spectrum than the one produced by these authors from only stable low/hard X-ray
intervals (Interval I in Rodriguez et al. 2008b), whereas the very short X-ray spikes originating from the disk 
(Intervals II and IV) are washed out. This shows that the SPI spectra of the
one day averaged comptonized emission are not influenced by any peculiar outburst-behavior on very short
time-scales. Likewise, a comparison with simultaneous Suzaku observations
leads to similar conclusions for observation 368.

\subsection{Cross-comparison}

When comparing the results of these four high-lighted observations, we notice that the X-ray variability and the
spectral behavior at higher energies ($>20\,keV$) seem to be completely uncorrelated.
Except for a slight difference in flux, the SPI spectra from observations
368 and 295 (illustrated in Figure 6) show the same characteristics,
whereas the temporal behavior in X-rays is seen to be dramatically different (compare Figure 4 left and Figure 5 right). A
similar high-energy correspondence can be observed between observations 246 and
423 (Figure 7), although again no similarities are found
in X-rays. We thus see that for similar X-ray variability, the
$20$\thinspace --\thinspace $500\,keV$ spectra can be different and reciprocally that for similar
(\thinspace $\sim $\thinspace one day averaged) high-energy spectral behavior, the
source may exhibit very unlike $1.2$\thinspace --\thinspace $12\,keV$ activity.

\begin{figure}[!t]
   \centering
   \includegraphics[width=\columnwidth]{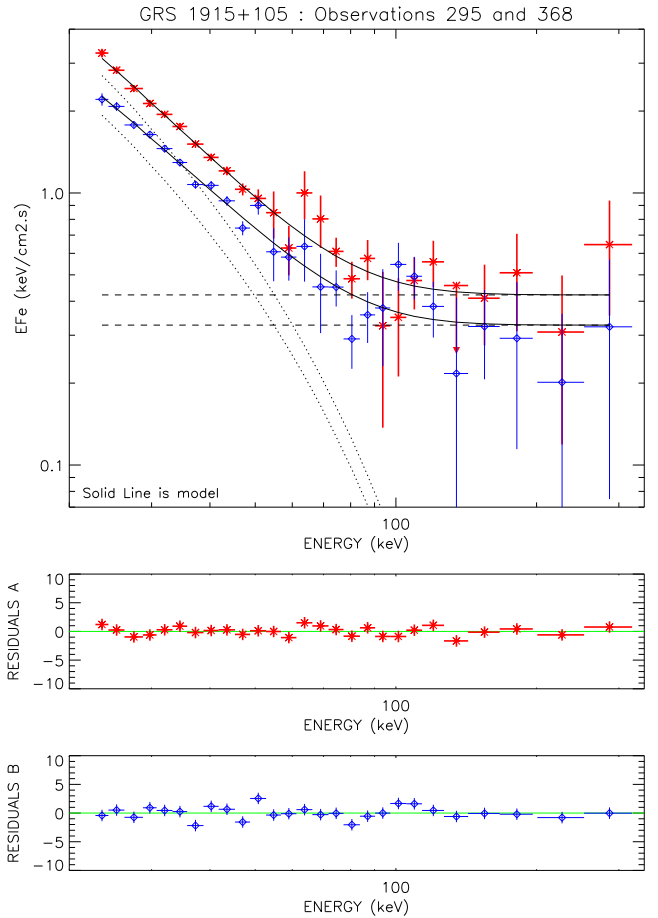}
      \caption{COMPTT+PL best fits for observations 295 (red stars) and 368 (blue diamonds). 
	Both observations have very similar spectral shapes.}
      \label{295368}
\end{figure}
\begin{figure}[!t]
   \centering
   \includegraphics[width=\columnwidth]{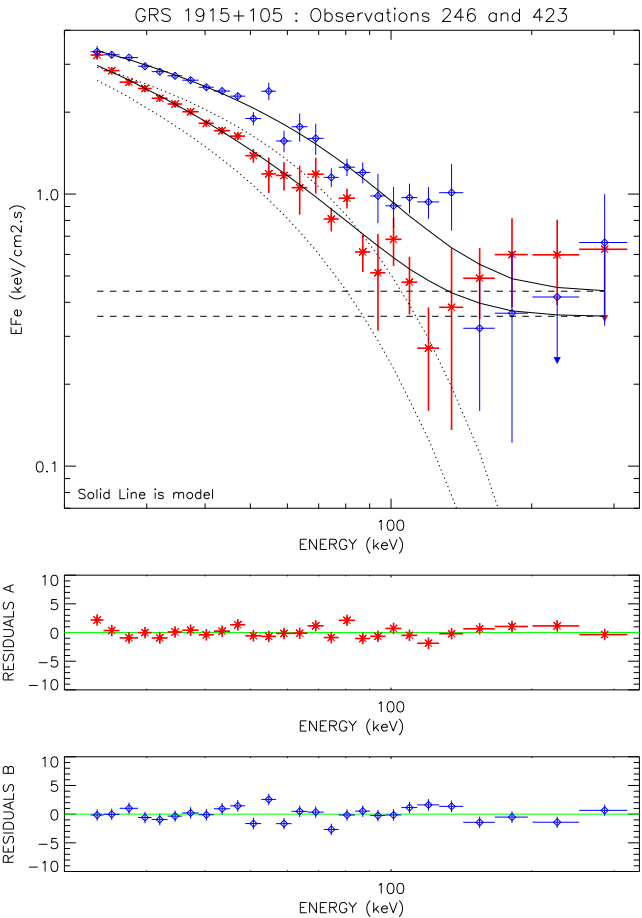}
      \caption{COMPTT+PL best fits for observations 246 (red stars) and 423 (blue diamonds).}
      \label{423246}
\end{figure}

\subsection{Composite spectra}

In order to further investigate the high-energy tail of GRS
1915+105, we decided to group data from observations with similar spectra. Improved statistics should
allow us to put better
constraints on the parameters of the fitted models. Hence, we isolated
two opposite groups in the spectral index versus flux diagram (Figure 3) for which we generated composite spectra. 
The first group
is characterized by a rather low $20$\thinspace --\thinspace $50\,keV$ flux ($%
\approx 200\,mCrab$) and a very soft spectral shape ($\Gamma \approx 3.45$);
hereafter we will call it the \textit{soft sample}. The second group on the
other hand has hard colors ($\Gamma \approx 2.90$) and high average flux ($\approx
330\,mCrab$) in the $20$\thinspace --\thinspace $50\,keV$ band; it will
accordingly be called the \textit{hard sample}. We note that the previously
presented observations 295/368 and 246/423 are representative of each group, respectively. As some observations had
intermediate characteristics, we
excluded them from the regrouped spectra in order to avoid mixing different high-energy patterns.
As a result, our two samples are likely to describe the boundary comptonization
states between which the source seems to be continuously switching.

We fitted both composite spectra shown in Figure 8 with several models and
summarize the results in Table 2. Each time the thermal\thinspace +\thinspace non-thermal comptonization models gave
the best fit to the data and demonstrated the need for an additional high-energy 
component (see \textit{F-TEST} values in Table 2). The composite spectral analysis thus confirms the previously
outlined interpretation, which will be further discussed in the next section.

\begin{table*}[ht]
\centering
\begin{tabular}{cccccccccc}
\hline
\T & & & & \multicolumn{2}{c}{SPI$_{\,22-50\,keV}$} & \multicolumn{2}{c}{ASM$_{\,1.2-12\,keV}$} & & \\
\B Obs ID & MJD start & MJD stop & Exp time $\scriptstyle{(s)}$ & $<\!F\!>\,\scriptstyle{(mCrab)}$ & $rms/\!\!<\!F\!>\,\scriptstyle{(\%)}$ & $<\!F\!>\,\scriptstyle{(cts/cm^{2}/s)}$ & $rms/\!\!<\!F\!>\,\scriptstyle{(\%)}$ & $\Gamma$ & $\chi ^{2}/27$\\
\hline\hline
\T 231 & 53253.20 & 53253.95 & 48412.50 & 322.5 $\pm $ %
5.0 & 20.2 & 51.9 $\pm $ 0.5 &  58.0 & 2.80 $\pm $ 0.06 & 0.97 \\ 
242 & 53285.10 & 53285.75 & 42782.20 & 259.5 $\pm $ 4.5 & 11.6 & 70.9 $\pm $ 0.7 & 8.20 & 3.09 $\pm $ 0.06 
 & 0.56 \\ 
243 & 53288.63 & 53289.44 & 49962.72 & 199.8 $\pm $ 4.4 & 16.0 & 62.1 $\pm $ 0.5 & 10.1 & 2.99 $\pm $ 0.10 
 & 1.08 \\ 
246 & 53296.35 & 53297.53 & 77174.90 & 305.8 $\pm $ 3.1 & 7.00 & 81.3 $\pm $ 0.3 & 28.2 & 3.02 $\pm $ 0.05 
 & 1.37 \\ 
248 & 53302.88 & 53304.90 & 134148.88 & 201.3 $\pm $ 2.7 & 18.5 & 69.4 $\pm $ 0.6 & 9.90 & 3.07 $\pm $ 0.07
 & 0.84 \\ 
250 & 53309.14 & 53310.22 & 72648.77 & 199.1 $\pm $ 3.9 & 18.0 & 71.7 $\pm $ 0.4 & 9.00 & 3.10 $\pm $ 0.10 
 & 0.71 \\ 
251 & 53311.32 & 53312.77 & 84373.42 & 320.4 $\pm $ 3.9 & 16.7 & 53.3 $\pm $ 0.6 & 48.4 & 3.00 $\pm $ 0.06
 & 1.12 \\ 
255 & 53324.27 & 53325.46 & 81422.62 & 152.1 $\pm $ 3.0 & 10.6 & no data & no data & 2.85 $\pm $ 0.09 
 & 0.64 \\ 
295 & 53442.95 & 53444.15 & 85681.48 & 255.5 $\pm $ 2.8 & 5.70 & 43.7 $\pm $ 0.4 & 5.10 & 3.47 $\pm $ 0.07
 & 1.40 \\ 
305 & 53472.66 & 53475.66 & 91334.75 & 180.2 $\pm $ 2.7 & 11.8 & no data & no data & 3.52 $\pm $ 0.08 
 & 0.96 \\ 
315 & 53503.53 & 53504.55 & 69072.34 & 158.3 $\pm $ 3.4 & 20.7 & 134.1 $\pm $ 0.7 & 26.6 & 3.00 $\pm $ 0.10
 & 1.26 \\ 
361 & 53640.92 & 53641.58 & 70136.04 & 123.2 $\pm $ 3.1 & 37.6 & 77.0 $\pm $ 0.7 & 35.7 & 3.24 $\pm $ 0.13
 & 0.92 \\ 
366 & 53655.52 & 53657.87 & 161541.02 & 169.7 $\pm $ 3.0 & 19.7 & 59.1 $\pm $ 0.6 & 9.00 & 3.47 $\pm $ 0.09 
 & 1.46 \\ 
367 & 53659.23 & 53660.83 & 109687.94 & 160.4 $\pm $ 3.6 & 17.1 & 108.2 $\pm $ 0.5 & 40.9 & 3.48 $\pm $ 0.11 
 & 1.41 \\ 
368 & 53661.51 & 53663.78 & 154549.29 & 187.2 $\pm $ 2.8 & 18.6 & 113.7 $\pm $ 0.4 & 28.7 & 3.42 $\pm $ 0.07 
 & 1.68 \\ 
369 & 53664.31 & 53665.53 & 52753.31 & 194.0 $\pm $ 6.4 & 24.7 & 116.9 $\pm $ 0.6 & 43.5 & 3.43 $\pm $ 0.14 
 & 1.06 \\ 
373 & 53676.27 & 53677.43 & 83798.38 & 87.2 $\pm $ 3.8 & 36.6 & 77.6 $\pm $ 0.7 & 38.9 & 3.14 $\pm $ 0.18
 & 1.07 \\ 
379 & 53694.27 & 53695.47 & 90421.32 & 183.1 $\pm $ 3.9 & 15.3 & 167.3 $\pm $ 1.5 & 32.6 & 3.08 $\pm $ 0.10 
 & 0.89 \\ 
423 & 53825.88 & 53828.23 & 119129.04 & 391.1 $\pm $ 4.2 & 7.90 & 29.1 $\pm $ 0.2 & 10.0 & 2.80 $\pm $ 0.04 
 & 3.07 \\ 
\B 431 & 53851.94 & 53852.20 & 15509.04 & 278.7 $\pm $ 7.2 & 7.30 & 26.6 $\pm $ 0.4 & 7.20 & 2.80 $\pm $ 0.10 
 & 1.46 \\ \hline\hline
\end{tabular}%
\caption{{\protect\small Log of the SPI observations and basic source
characteristics. Temporal behavior is parameterized in terms of average flux and fractional rms whereas the photon index $\Gamma $ and $\chi ^{2}/\nu $
are the parameter and quality of a simple powerlaw model fit.}}
\label{tab1}
\end{table*}

\begin{table*}[ht]
\centering%
\begin{tabular}{ccccccccccc}
\hline
\T & \multicolumn{6}{c}{ COMPTT / COMPPS} & \multicolumn{2}{c}{PL} &  &  \\ 
\BB Obs ID & $kT_{bb}$ {\ (keV)} & $kT_{e}$ {(keV)} & $\tau $ & $\gamma _{\min }$
& $\Gamma _{e}$ & F$_{comptt}$ & $\alpha $ & K$_{pl}$ & $\chi ^{2}/\nu $ & 
FTEST \\ \hline\hline
\T \B &  & {83}$_{-2.3}^{+4.2}$ & {0.08} & - & - & - & - & - & 32/26 &  \\ 
\B 246 & 1.0f & {18.2}$_{-1.0}^{+1.1}$ & {0.97} & - & - & 3.64$_{-0.04}^{+0.04}$ & 2.0f & 3.48$_{-0.96}^{+1.05}$ & 
27/25 & 4$\times $10$^{-2}$ \\ 
\B &  & 16.20$_{-0.26}^{+0.31}$ & 3.68$_{-0.20}^{+0.23}$ & 1.36 & 2.12 & - & -
& - & 28/24 &  \\ 
\TT \B &  & {44} & {0.18} & - & - & - & - & - & 72/26 &  \\ 
\B 295 & 1.5f & {16.3}$_{-0.9}^{+1.2}$ & {0.57} & - & - & 2.73$_{-0.03}^{+0.03}$ & 2.0f & 3.95$_{-0.56}^{+0.66}$
& 17/25 & 2$\times $10$^{-9}$ \\ 
\B &  & 14.43$_{-0.23}^{+0.23}$ & 2.70$_{-0.17}^{+0.12}$ & 1.30 & 2.36 & - & -
& - & 17/24 &  \\ 
\TT \B &  & {39} & {0.20} & - & - & - & - & - & 74/26 &  \\ 
\B 368 & 1.5f & {13.4}$_{-0.9}^{+1.0}$ & {0.86} & - & - & 1.94$_{-0.03}^{+0.03}$ & 2.0f & 3.28$_{-0.56}^{+0.54}$ & 
29/25 & 1$\times $10$^{-6}$ \\ 
\B &  & 15.10$_{-0.29}^{+0.27}$ & 2.64$_{-0.20}^{+0.21}$ & 1.31 & 2.35 & - & -
& - & 29/24 &  \\ 
\TT \B &  & {41}$_{-1.7}^{+1.2}$ & {0.45} & - & - & - & - & - & 42/26 &  \\ 
\B 423 & 1.0f & {19.2}$_{-1.0}^{+1.0}$ & {1.20} & - & - & 4.91$_{-0.05}^{+0.05}$ & 2.0f & 4.27$_{-1.38}^{+1.29}$ & 
37/25 & 8$\times $10$^{-2}$ \\ 
\B &  & 17.54$_{-0.28}^{+0.31}$ & 4.40$_{-0.20}^{+0.20}$ & 1.38 & 2.11 & - & -
& - & 37/24 &  \\ 
\TT \B &  & {48} & {0.13} & - & - & - & - & - & 152/26 &  \\ 
\B SS & 1.5f & {16.5}$_{-0.6}^{+0.6}$ & {0.62} & - & - & 2.07$_{-0.03}^{+0.03}$ & 2.0f & 2.75$_{-0.29}^{+0.28}$ & 
31/25 & 3$\times ${10}$^{-10}$ \\ 
\B &  & 14.72$_{-0.13}^{+0.16}$ & 2.70$_{-0.25}^{+0.10}$ & 1.31 & 2.35 & - & -
& - & 31/24 &  \\ 
\TT \B &  & {38}$_{-0.9}^{+0.9}$ & {0.47} & - & - & - & - & - & 30/26 &  \\ 
\B HS & 1.0f & {17.7}$_{-0.7}^{+0.7}$ & {1.28} & - & - & 4.13$_{-0.04}^{+0.04}$ & 2.0f & 3.31$_{-0.95}^{+0.91}$ & 
19/25 & 8$\times ${10}$^{-4}$ \\ 
\BB &  & 18.14$_{-0.37}^{+0.31}$ & 4.02$_{-0.14}^{+0.15}$ & 1.39 & 2.21 & - & -
& - & 19/24 &  \\ \hline\hline
\end{tabular}%
\caption{{\protect\small Spectral fitting results for the high-lighted observations. For each spectrum, the first row
corresponds to a simple thermal comptonization model (COMPTT), the second row contains the COMPTT
+ PL parameters and the last gives the variables of the COMPPS model. The seed photon temperature is fixed to $1.0\,keV$ and 
$1.5\,keV$ for the harder and softer spectra respectively. 
COMPTT errors are calculated on $kT_{e}$ while fixing $\tau$ to its best fit value.
For the COMPPS model, the geometry parameter is fixed to zero (spherical) and neither reflection
nor ionization is taken into account. The error bars are simultaneously evaluated on $kT_{e}$ and
$\tau$ while we fixed $\gamma _{\min }$ and $\Gamma _{e}$ to their best fit values. 
F$_{comptt}$ denotes the integrated 20-500 keV energy flux from the thermal component and is given in
units of $\times 10^{-9}erg/cm^{2}/s$. K$_{pl}$ is the flux normalization at $100\,keV$ of the
non-thermal power law and is given in units of $\times 10^{-5}photons cm^{-2} s^{-1} keV^{-1}$}}
\label{tab2}
\end{table*}

\section{Discussion}

We interpret the $20$\thinspace --\thinspace $500\,keV$ emission of GRS
1915+105 as a combination of thermal and non-thermal comptonization. Given
the similarities of the observed temporal properties, the ASM band is likely
to sample the same emission component as the $20$\thinspace --\thinspace $%
50\,keV$ SPI band. This is in agreement with the results of Done, 
Wardzi\'nski \& Gierli\'nski (2004) who show that except for ultrashort disc-dominated 
X-ray spikes the accretion disc has no significant effect above $3\,keV$ (see also Rodriguez et al. 2008b). 
The spectra obtained during observations 295 and 368 show that for different X-ray classes
(i.e. different temporal behavior of the accretion flow) the high energy spectra can be very similar. 
Conversely, observations 295
and 423 show that within stable emission episodes, there can be significant
differences in the spectral behavior of the source. 
Assuming that the $3$\thinspace --\thinspace $50\,keV$ emission originates from the comptonizing corona
(see Malzac 2007 or Done et al. 2007 for details), this clearly indicates
that there is no correlation between the temporal variability and the
macroscopic properties of the comptonizing flow. As they give better statistics above $100\,keV
$, we will use our composite spectra to further discuss both
comptonization components and put our findings in a more general context.

\subsection{The soft sample}

From spectral modelling we find the intersection point of the two components
to be around $60\,keV$. At higher energies, the non-thermal processes dominate. Both the thermal
and non-thermal components have approximately the same luminosity above $20\,keV$. 
If these two components originate from the
same population (COMPPS results), the electrons are found to thermalize around Lorentz factor 
$\gamma_{min}$\thinspace$\approx$\thinspace $1.30$. The mildly hot Compton cloud 
($kT_{e}=14.72\pm 0.15\,keV$) 
is found to be marginally optically thick ($\tau=2.7_{-0.25}^{+0.10}$)
which is the expected configuration for black-hole binaries in the very high state (Done \& Kubota 2006).
We suggest that our soft composite spectrum gives a canonical high-energy representation of GRS
1915+105 in low coronal luminosity states.

\subsection{The hard sample}

Concerning the hard sample, the luminosity of the non-thermal component is found to
be roughly the same (compatible within the uncertainties) as for the previously discussed soft sample. However, the intersection 
point of the two components is now around $110\,keV$ (which corresponds to a Lorentz factor $\gamma_{min}$\thinspace
$\approx$\thinspace $1.39$), showing that the main difference lies in the properties of
the comptonizing thermal electrons of the corona. The plasma is found to be
either hotter for similar optical depths or optically thicker for similar
electron temperatures (or a mixture of both), thus enhancing the higher observed $20$\thinspace
--\thinspace $50\,keV$ flux. Even though this situation cannot be completely
resolved due to the observational $kT$\thinspace --\thinspace $\tau $
degeneracy, spectral fits with COMPPS indicate that it is likely that there has been
a significant increase in opacity, whereas electron temperature remains around $15 - 20\,keV$. 
In any case, this does not affect the
estimation of the total $20$\thinspace --\thinspace $500\,keV$ luminosity
issued from thermal Compton scattering, which is found to be enhanced by a factor of $2\pm 0.04$ in comparison with the soft sample.
We interpret our spectrum as
a typical representation for the high-energy emission of GRS 1915+105 in high
coronal luminosity states.

\begin{figure}
   \centering
   \includegraphics[width=\columnwidth]{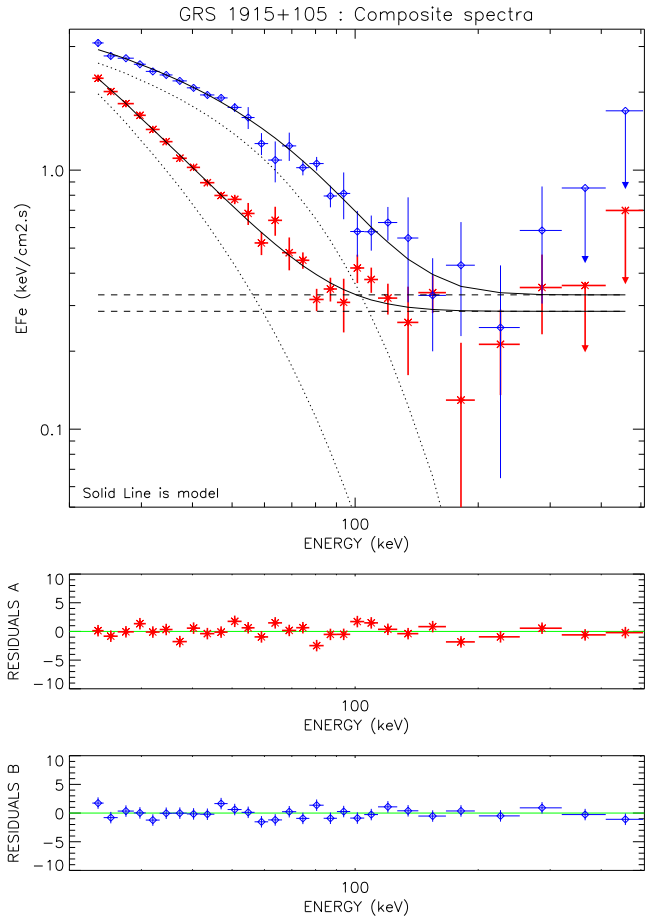}
      \caption{COMPTT+PL best fits for the composite spectra, plotted in red stars and blue diamonds for the soft and hard samples respectively.
	These spectra illustrate the boundary characteristics of the one day averaged high-energy emission of GRS1915+105.}
      \label{composite}
\end{figure}

\section{Summary and Conclusions}

We have conducted detailed high-energy spectral analysis of the microquasar GRS 1915+105
using all available SPI data from July 2004 to May 2006. We
presented here our observational results on the source's high-energy output. We can
summarize our findings as follows:

We found that the $\sim$\thinspace one day averaged $20$\thinspace --\thinspace $500\,keV$ spectral
emission is always between two boundary states, \textit{hard} and \textit{%
soft}, which we illustrated through spectral modelling.
We confirm that in the INTEGRAL-SPI data we observe no high-energy cutoff for GRS 1915+105
(Fuchs et al. 2003, Rodriguez et al. 2008b). We suggest that
the high-energy cutoff from thermal comptonization is drowned out by an
additional non-thermal component. We found the non-thermal component to be
statistically required in both composite samples. The spectral differences we observed in
hard X-rays ($20$\thinspace --\thinspace $50\,keV$) are most likely to be coupled to the evolution of the thermal electron plasma.
The bolometric luminosity calculated from the thermal
component varies by a factor of $2$ between soft and
hard samples. In contrast, the obtained fits indicate that the non-thermal
component is fairly stable. This implies that both components are not
necessarily linked, i.e. they could originate from dissociated electron
populations. Hence as suggested by Rodriguez et al. (2008b), the non-thermal
component might be from emission from the jet.
Given the length of the high-energy observations (SPI\thinspace $\approx$\thinspace $3\,days$ or OSSE\thinspace $\approx$\thinspace $15\,days$), it is
difficult to investigate the connection between the various X-ray classes and
the high-energy spectra. We pointed out that there is no direct correlation between the observed X-ray variability patterns and our 
$20$\thinspace --\thinspace $500\,keV$ SPI spectra.
This shows that the macroscopic properties of the comptonizing thermal electrons 
evolve independently from the temporal behavior of the source, i.e. independently from the fluctuations of the accretion flow. 
These aspects will be adressed in a subsequent paper using multiwavelength coverage.
In summary, our results give a recent high-energy picture of GRS 1915+105 and once more underline the complex nature of the 
accretion processes operating around this archetypal microquasar.

\begin{acknowledgements}
The SPI project has been completed under the responsibility
and leadership of the CNES. We are grateful to the ASI, CEA, DLR,
ESA, INTA, NASA, and OSTC for support.
Specific software packages used for this work have been developed
by E. Durand at the CESR, Toulouse. We are grateful to the anonymous
referee for the fruitful comments that allowed us to
improve the quality of this paper.
\end{acknowledgements}

\end{document}